\documentclass[12pt]{article}
\usepackage{graphics,epsfig}
\usepackage[english]{babel}
\begin{document}


\begingroup

\raisebox{0.5cm}[0cm][0cm] {
\begin{tabular*}{\hsize}{@{\hspace*{5mm}}ll@{\extracolsep{\fill}}r@{}}
\begin{minipage}[t]{3cm}
\vglue.5cm
\end{minipage}
&
\begin{minipage}[t]{7cm}
\vglue.5cm
\end{minipage}
&
\begin{minipage}[t]{7cm}

\end{minipage}
\end{tabular*}
}

\begin{center}

{\Large{\bf Self-induced radiation of 4.3 GeV electron \\beam in
crystalline medium}}

\vspace{1.cm}
                {\bf A.Aganyants}

\setlength{\parskip}{0mm}
\small

\vspace{1.cm}
              {\bf Yerevan Physics Institute, Armenia}
\end{center}

\vspace{1.cm}

\begin{abstract}

Investigations of interactions of a relativistic electron beam
with single crystals have shown that yields of gamma rays, their
energies and outgoing angles grow non-linearly with the beam
intensity especially at small incidence angles of electrons
relative to crystalline planes and axes. It is observed peculiar
role of a crystal boundary in these phenomena. One can consider
coherent properties of physical vacuum, manifested near the
crystal boundary are responsible for correlation processes in
electromagnetic interactions of the relativistic electrons.

\end{abstract}


\newpage
\section{Introduction}

  As known non-equilibrium processes perform constructive role in physics,
  chemistry and biology. They define origin coherent structures and
  self-organization process \cite{ref1}. The stronger non-equilibrium the deeper
  inter -connection of different phenomena and essences.
   That is way any more energy, laser fields, currents
    and intensity of irradiation is used. Lee \cite{ref2}, for example, proposed
    to investigate the coherent properties of physical vacuum
    exciting it's by colliding ultra-high energy heavy nuclei.\\
   The same concerns electromagnetic vacuum. Quantum electrodynamics
   is studied well on short distances. Electrodynamics of
   intense fields is needed in experimental investigations. As will be shown below growth
    of intensity of the electron beam result in manifesting the coherent properties of
    vacuum due to which correlation of excited atoms comes, responsible for
   modifications of electromagnetic processes in crystalline medium.

\section{Experimental results}

  Experiments carried out by means of 4.3 GeV electrons moving in single crystals
   of  diamond and others since 1976 on the internal electron beam of
   Yerevan Synchrotron. The advantage of the last is multiple passage of
   electrons through thin target, that equivalently acting more intense
   current of accelerator. Evidence in paper \cite{ref3} indicates that
   crystalline medium excited by intense electron beam influences on
electromagnetic cross-sections of passing relativistic electrons.
Data presented in paper \cite{ref3}  show anomalous growth of
emission angles of gamma rays. New data, involved in this paper,
concern effects of crystal edges in conditions when size of
crystal is commensurable with beam one or electron beam pass
through crystal near its edge. These conditions, which take place
in all cited papers, apparently, are responsible for non-linear
processes, mentioned in the beginning of this article and other
\cite{ref4,ref5}.
  In this paper new experimental data are presented together with some published
   data for confident evidence of afro-cited statements.
  Preliminary investigations of interactions of electron bunches passing near
  planes (110) and axes [100] have shown:

  1. Arising intense radiation is low-energy one in main. This conclusion
   based on such a fact. Synchrotron radiation losses of electrons in
   accelerator are $\approx$ 1.5 MeV per one revolution. If energy losses in
   crystal are compatible with this value then electrons at defined
   tuning of the accelerator can dump to the vacuum chamber wall near the
    diamond target in main. The experimental setup is presented in \cite{ref3}.
     Detector situated near the target really showed sharp growth of
     counting rate in depending on beam intensity and crystal orientation \cite{ref4}.
      However counting rate of this detector significantly exceeds one of
       any gamma-quanta detectors. It is one of evidence that a part of
   anomalous radiation extends large angles and therefore does not
   pass through collimator or has very low energy.

  2. Anomalous radiation arises when intensity of the electron beam passing
   through single crystal exceeds some threshold, i.e. radiation in a
   crystal increases non-linearly. Integrated yield of gamma rays grows
   by power law with the electron beam intensity up to $I^{4}$
   (evidence of a phase transition), where I is electron current.
   In doing so it is observed also strong almost 100\%
   dumping of accelerated electrons when their bunches pass under the
   small angles relative to planes (110) of the crystal and  just only
   4\% at the disoriented crystal.

  3. Production of intense gamma rays is observed with the diamond
  crystal of 100 and 470 $\mu$m. Yield of low-energy gamma quanta from
  the crystal (470 $\mu$m) approximately equal to one with thickness 100 $\mu$m.
  It means production of soft radiation occurs due to interaction in
   front part of the crystal. That is a consequence of large interaction
   cross-section of electron and subsequent screening of other atoms \cite{ref4}.

   4. Degree of gamma beam polarization is high, when electrons travel
   through the crystal in parallel to planes (110). This conclusion was
   made on the base of measured azimuth asymmetry 0.81 \cite{ref3}.
   This asymmetry was measured at scattering of low energy radiation
   by amorphous target (polystyrene of 1.5 cm thick) under the angle of 20 mrad.
    Azimuth asymmetry of linear Compton- scattering of hard gamma quanta at
    small angles does not exist. One can think observed azimuth asymmetry
    is connected with property of boundary of excited polystyrene
     medium and with high coherence of gamma beam.

  5. As noticed in \cite{ref3} quantities of total radiation passing
  through two different collimators non-proportional to the ratio of their
  aperture areas. Possible explanation is given below.
  New similar results were obtained in conditions when electron beam
  passed through edge of diamond target. Produced gamma rays traveled
  through collimator with subsequent scattering of photons under the
   angle of 3.4$^{\circ}$ relative to the direction of gamma beam and are
   detected by means of simple counter with thin scintillator of ZsJ.
    The energy of accelerated electrons was 2.9 GeV and scattering
    target was polystyrene of 5mm thickness, situated on the distance
    of 28 m from diamond-radiator. A strong magnetic field swept all
     produced charged particles. It is studied dependence of detector
      counting rates on crystal orientation firstly when radiation from
    diamond passes through collimator aperture 6.7$\times$6.7 mm$^2$ and then
    3.3$\times$3.3 mm$^2$. Measured counting rates were normalized
    to the quantameter data. The normalized counting rates sharply
    fell down at orientation of crystal plane (110) near 0$^{\circ}$
     in case of collimator aperture 3.3$\times$3.3 mm$^2$ (table 1).

\begin{table}[ht]
\begin{center}
\begin{tabular}{|c|c|c|c|}
  \hline

Orientation, mrad&Aperture 6.7$\times$6.7&Aperture 3.3$\times$3.3
\\ \hline
0.0&1665000&587000\\ \hline 0.085&1660000&597000\\ \hline
0.42&1430000&590000\\ \hline0.85&1580000&627000\\ \hline

\end{tabular}

\end{center}
\caption{Normalized counting rates of ZsJ detector versus the
collimator aperture and crystal orientation.}
\end{table}

So cutting off the normalized yields of radiation almost 3 times
is observed. One can think that this phenomenon arises because of
a part of intense radiation is produced on the edge of radiator
and has more wide angular distribution and different spectrum. In
item 6 last argument (edge) for that conclusion is demonstrated
more brightly.

  6. In this measurements movable diamond radiator with transversal
  size of 2 mm could displace together with electron beam in
  transversal direction relative to aperture of collimator 3.3$\times$3.3 mm$^2$,
   which was placed on the distance of 9.4 m from radiator.
   So collimator took gamma beam only from defined part of diamond.
   Monitoring of total energy of radiation was carried out by means of
   Wilson quantameter. Table 2 present the counting rates of the detector,
   situated under angle 6$^{\circ}$ relative to gamma beam direction and the same
   detector under angle 19$^{\circ}$  versus coordinate of diamond displacement r.
   Crystal orientation relative to planes (110) was 0$^{\circ}$ . It is seen increase
    of the detector counting rate when collimator  "sees" only an edge of
    the crystal. Why the edge?

\begin{table}[ht]
\begin{center}
\begin{tabular}{|c|c|c|c|}
  \hline

r, mm&Detector 6$^{\circ}$&Detector 19$^{\circ}$
\\ \hline
0&16000&995\\ \hline 0.5&25100&1300\\ \hline 2&38400&1920\\ \hline

\end{tabular}

\end{center}
\caption{Normalized counting rates of detectors depending on the
cristal displacement relative to the collimator aperture.}
\end{table}

Breaking spatial symmetry near edges of crystal \cite{ref6} is
simple explanation. In this topology vacuum is polarized. Coherent
properties of vacuum result in ordering atomic excitations due to
primary polarization, i.e. synchronization of atomic fields, which
create a strong field. Relativistic electrons moving in that field
produce intense synchrotron-like radiation. Apparently, broadened
angular distributions of photons with energies 300-800 MeV in
\cite{ref7} demonstrate this statement. So symmetry breaking on
the boundary is responsible for arising strong field on the edges
of excited crystal.
  There is experimental evidence, that coherent bremsstrahlung and
  channeling one in the same conditions are decreased.  Cause for such
   a behavior is absence of particle oscillations on its way because of
    superposition of mentioned low-frequency strong field. Electron will
    be deflected in some direction on the more large distance already.
     Respectively angles of gamma-emission grow also.

  7. There are measurements, which show that even a very hard part
  of bremsstrahlung spectrum grows with electron beam intensity at
  any orientation of the crystal. That growth more two times is
  shown in table 3. Particles are detected in this experiment only
   from hadrons produced in the reaction $\gamma$ + nucleus $\rightarrow$ hadrons, as
   detector was placed under large angle of 19$^{\circ}$ relative
    to gamma beam direction. Electron beam intensity was increased
     in this measurement due to faster beam-dump.

\begin{table}[ht]
\begin{center}
\begin{tabular}{|c|c|c|c|}
  \hline

Orientation, mrad& Slow beam-dump & Faster beam-dump
\\ \hline
0.0&551&1180\\ \hline 0.34&703&1630\\ \hline 1.0&894&2110\\ \hline

\end{tabular}

\end{center}
\caption{Normalized counting rates of detector versus electron
beam intensity and crystal orientation.}
\end{table}

These measurements showed strengthening of the atomic fields at
small impact parameter and prove once again existence of mentioned
strong field because of atomic correlation.

\section{Conclusion}

  So obtained data in item 1-7 show that radiation cross-sections of
   relativistic electrons, the energy of emitted photons and outgoing
    angles of hard radiation grow by virtue of medium coherence when
     intensity of the electron beam is increased.
  Process of ordering of atomic interaction result in both different
 kinds of photon emission of relativistic electrons, which are
 superposed against each other in radiation spectrum:\\
  1. Cherenkov, quasi-Cherenkov,  transition radiation.
  They have polarization nature, when high-energy electron interacts with
  atoms.\\
  2.Bremsstrahlung, coherent bremsstrahlung, channeling and synchrotron-like
   radiation , when electron does with atomic Coulomb-fields of oriented
   crystal.\\
  When crystal is excited all these processes are modified. Coherent
  bremsstrahlung and channeling radiation are partially suppressed,
   however transition and synchrotron-like radiation are
   strengthened.\\
  Different applications are possible for described phenomena
  It is possible creation (in dependence on energy of electrons,
  their beam intensity and kind of accelerators):\\
1.Intense source of linear polarized gamma quanta, in particular,
gamma-laser.\\ 2.Method for measure the degree of gamma beam
linear polarization.\\ 3.Intense source of positrons.\\
4.Quark-gluon plasma, if to collide intense gamma beams, produced
on crystals from counter-electron beams of future linear
colliders. \\In measurements, marked by item 5 Garibyan V. and
Vartanov Yu. participated also. The work is supported by contract
116 of RA government and partially by CRDF AP2-2305-YE-02.


\end{document}